\documentclass[11pt]{article}

\usepackage[skip=10pt plus1pt, indent=15pt]{parskip}
\usepackage{amsmath}
\usepackage{amssymb}
\usepackage{mathtools}
\usepackage{amsthm}
\usepackage{graphicx}

\usepackage[english]{babel}
\usepackage{fancyhdr}
\usepackage{color}
\usepackage{bbm}
\usepackage{bm}
\usepackage{natbib}
\usepackage{wrapfig}
\usepackage{algorithmicx}
\usepackage{algorithm}
\usepackage{booktabs}
\usepackage[noend]{algpseudocode}
\usepackage[font={small}]{caption}
\usepackage{float}
\usepackage[section]{placeins}
\usepackage{sidecap}
\usepackage{tikz}
\usetikzlibrary{decorations.markings}
\usepackage{comment}

\usepackage[margin=1in]{geometry}

\makeatletter
\def\BState{\State\hskip-\ALG@thistlm}
\makeatother

\newcommand{\E}{\mathbb{E}}

\newcommand\bX{\bm{X}}
\newcommand\balp{\bm\alpha}
\newcommand\bbet{\bm\beta}

\usepackage{hyperref}
\usepackage[margin=1in]{geometry}
\hypersetup{colorlinks,citecolor=blue,urlcolor=blue,linkcolor=blue}

\newtheorem*{lemma*}{Lemma}

\theoremstyle{definition}

\title{Causal Estimation of Position Bias in Recommender Systems Using Marketplace Instruments}

\author{
Rina Friedberg\footnote{RF and KR contributed equally to this work.} \\ rfriedberg@linkedin.com \and 
Karthik Rajkumar$^{*}$ \\ krajkumar@linkedin.com \and 
Jialiang Mao \\ jimao@linkedin.com \and 
Qian Yao \\ qiyao@linkedin.com \and 
YinYin Yu \\ yinyyu@linkedin.com \and 
Min Liu \\ mliu@linkedin.com
\\ 
}

\begin{document}
\maketitle 

\begin{abstract}
Information retrieval systems, such as online marketplaces, news feeds, and search engines, are ubiquitous in today's digital society. They facilitate information discovery by ranking retrieved items on predicted relevance, i.e. likelihood of interaction (click, share) between users and items. Typically modeled using past interactions, such rankings have a major drawback: interaction depends on the attention items receive. A highly-relevant item placed outside a user's attention could receive little interaction. This discrepancy between observed interaction and true relevance is termed the position bias or effect. Position bias degrades relevance estimation and when it compounds over time, it can silo users into false relevant items, causing marketplace inefficiencies. Position bias may be identified with randomized experiments, but such an approach can be prohibitive in cost and feasibility. Past research has also suggested propensity score methods, which do not adequately address unobserved confounding; and regression discontinuity designs, which have poor external validity. In this work, we address these concerns by leveraging the abundance of A/B tests in ranking algorithm evaluations as instrumental variables. Historical A/B tests allow us to access exogenous variation in rankings without manually introducing them, and harming user experience and platform revenue. We demonstrate our methodology in two distinct applications at LinkedIn — feed ads and the People-You-May-Know (PYMK) recommender. The marketplaces comprise users and campaigns on the ads side, and invite senders and recipients on PYMK. By leveraging prior experimentation, we obtain quasi-experimental variation in item rankings that is orthogonal to user relevance. Our method provides robust position effect estimates that handle unobserved confounding well, greater generalizability, and easily extends to other information retrieval systems. 

\end{abstract}

\section{Introduction}
Information retrieval systems are ubiquitous in today's digital society: online marketplaces, social networks, search engines and content recommender platforms are just a few examples. Among their principal functions is to optimize the efficiency and quality of information discovery. Frequently, they accomplish this goal by ranking the retrieved items by predicted relevance between the user and an item, defined as the expected probability of interaction (e.g. click, share, purchase) conditional on the item receiving adequate user attention. These relevance models are fit using past observed interactions between users and items as training labels. This modeling process has a major drawback, however, in that the likelihood of interaction is not only a function of the item's true relevance, but also of the attention it has received. An item placed in a region outside of the user's attention could receive little interaction even if it was highly relevant. This discrepancy between the observed likelihood of interaction and the true relevance is termed the position bias or position effect \citep*{10.1145/1341531.1341545}. A consequence of the position effect is that training relevance predictive models based on observed user-item interaction could be severely biased, where an item getting randomly placed higher on the ranked list receives an unfair advantage, a discrepancy that could compound over each iteration of model update. This bias not only degrades item relevance estimation, but the compounding effect of position bias could also silo users into false relevant items with serious implications for platforms. In online marketplaces for ads, for example, because the amount charged for ads is proportional to its relevance \citep*{35261}, biased relevance estimate leads to overcharging advertisers or platform revenue loss. An accurate estimate of the position bias is thus extremely useful for training appropriately debiased ranking models. 

Solutions in the literature to estimate position bias broadly fall into three categories:
\begin{enumerate}
    \item Introducing random variations in position: techniques range from adding noise to predicted relevance or relevance ranking \citep*{conf/vldb/PandeyROCC05}, randomly swapping two items \citep*{10.1145/1341531.1341545, 10.1145/3018661.3018699}, to completely reshuffling the order of ranked items. While manually introducing random variation is the gold standard for estimating position effect, serving randomly ranked items is not optimal for users (who are served irrelevant content), items (lowered demand from misallocation) or the platform (steep revenue loss). For these reasons, this approach is generally confined to a small fraction of user traffic \citep*{10.1145/1341531.1341545}, which limits the statistical power in detecting position effects. An alternative is to limit the negative impact of rank-swapping intervention compared to uniformly randomizing rankings \citep*{10.1145/3018661.3018699}. 

    \item Estimating position bias with inverse propensity weighting \citep*{10.1145/3289600.3291017}: propensity score based methods have minimal user, item and platform impact as they can operate on observational data. However, these methods rely on strong assumptions, such as no uncontrolled confounders and a correctly-specified propensity model. In both of our worked examples, most past ranking functions changed due to improved relevance predictions, and hence rank qualitatively different search results in different locations. That is, we cannot rely on propensity score based methods to address unobserved confounding, especially through the important channel of item relevance.
    
    \item Regression discontinuity designs (RDDs) \citep*{doi:10.1287/mksc.2014.0893}: RDDs can only estimate position bias where the predicted relevance scores of two adjacent items are very close. Such item pairs are correlated with low item relevance or low quality relevance models which do not effectively model the relevance difference between two similar items. For this reason, RDDs typically do not have external validity when generalized beyond low-relevance items or platforms with ineffective relevance models.
    
\end{enumerate}

A complementary body of literature builds on these methods and estimate position bias with further modeling, such as by taking advantage of Bandit feedback \citep*{10.5555/3045118.3045206}, structural assumptions  \citep*{Li2020HandlingPB}, expectation-maximization models \citep*{46485}.

Our approach addresses all aforementioned shortcomings by leveraging quasi-experimental variations in rank positions introduced by historical A/B tests. Such tests are abundant in online information retrieval systems, as they are the standard process for evaluating new relevance models and product features. In the LinkedIn ads marketplaces alone, at least 50 such variations are introduced every year. Leveraging historical A/B tests allows us to access exogenous variations without manually introducing them and hurting user experience or platform revenue. The impact of ranking changes, as induced by these exogenous variations, is then estimated through carefully designed instrumental variable analyses. Specifically, our methodology exploits the two-sided nature of modern online marketplaces, using experimentation on the ranking of items in a manner orthogonal to user-side relevance. To our knowledge, this is the first paper to use such a marketplace instrument to estimate position bias in online ranking. We demonstrate its applicability in two distinct marketplaces on LinkedIn --- ads marketplace and the People-You-May-Know (PYMK) connections recommendation engine. Our empirical results show robust position effects across both the ads and PYMK marketplaces. By virtue of the A/B test instruments, our methodology handles unobserved confounding well. It also has greater external validity since the relevance model variation we utilize affects all users and all items, with an added advantage being the much larger data in this design compared to RDDs or manual experimentation. Finally, our methodology is generic and easily extends to all information retrieval systems.

The remainder of the paper is organized as follows. Section \ref{methodology} formally defines the position bias estimation problem and outlines the proposed solution. Sections \ref{sec:ad-application} and \ref{sec:pymk-application} share applications of the proposed methodology in two LinkedIn marketplaces: ads marketplace and the People-You-May-Know (PYMK) product. Section \ref{discus} discusses insights gleaned from position effects, provides future research directions and concludes.

\section{Methodology}
\label{methodology}
In this section, we lay out the instrumental variables framework through which we study causal position effects. 
\subsection{Notations and the causal estimands}

We have a group of users (viewers of feed or PYMK), who interact with or respond to a set of items (ads or candidates for PYMK) under certain context, where the response event can be defined as a click, a conversion, or other deep funnel feedback from the user. We seek to study the effect of an item's position changing on the user's propensity to interact with it, all else held equal. 

Let $I$ and $J$ denote items and requests and let $i$ and $j$ be a specific item and request, respectively. For each request, multiple items are shown to the user based on the underlying serving system with various relevance models. In request $j$, let $e_{ijk}=\mathbbm{1}(\text{response}_{ijk})$ be the counterfactual response indicator given that item $i$ is shown in position $k$ in request $j$. In this work, we only consider binary responses. We define the item-level and system-level position effect between any positions $k_1>k_2$ as
\begin{equation}
\begin{aligned}
\tau_{i}(k_1, k_2) & = \mathbbm{E}_J[e_{ijk_1} \mid I = i] -  \mathbbm{E}_J[e_{ijk_2} \mid I = i], \\
\tau(k_1, k_2) & = \mathbbm{E}_I[\tau_{i}(k_1, k_2) ].
\end{aligned}
\label{eq:poseffdef}
\end{equation}
These position effects could be interpreted as the expected change in users' response once an item is moved from position $k_2$ to $k_1$. In this work, our goal is to estimate these position effects in two use cases we face at LinkedIn. Although the contexts are specific, the methodology proposed to identify and estimate these effects are general and can be applied to other use cases. For example, one could also focus on the request-level position effects defined as $\tau_{j}(k_1, k_2) = \mathbbm{E}_I[e_{ijk_1} \mid J = j] -  \mathbbm{E}_I[e_{ijk_2} \mid J = j].$ 

To estimate $\tau_{i}$ and $\tau$, we rely on observations at the request level. Specifically, each observation corresponds to a (item, request, position) tuple.  Table \ref{tab:ex} shows some example observations in this format. 

\begin{table}[H]
\centering
\begin{tabular}{@{}c | ccc@{}}
\toprule
Response & Item & Request & Position \\ \midrule
1        & 1       & 1       & 1        \\
0        & 2       & 1       & 2        \\
0        & 3       & 1       & 3        \\ \midrule
0        & 1       & 2       & 1        \\
1        & 3      & 2       & 2        \\
0        & 6       & 2       & 3        \\ \midrule
$\vdots$         &    $\vdots$        &    $\vdots$        &   $\vdots$          \\ \bottomrule
\end{tabular}
\caption{Example observations.}
\label{tab:ex}
\end{table}

The key challenge in estimating the position effects with such observational data is that the responses are confounded by factors such as the relevance. We address confounding using an instrumental variables approach. 

\subsection{Instrumental Variables Model}

We leverage instrumental variables \citep*{10.2307/2291629} provided by past controlled experiments in ranking algorithms to estimate the position effect. 

We first focus on a specific item $i$ and consider the item-level position effect $\tau_i(k_1, k_2)$. Suppose that there was a past request-side experiment that affected the items' position, and only affected users' response through its impact on the positions. Consider all requests in which item $i$ was shown either in position $k_1$ or $k_2$ during the experiment. Let $W_j$ be the position of the item. Let $Z_j$ be the treatment request $j$ received in the selected experiment and let $Y_j$ be the corresponding binary response (click, connection request etc.). We consider the following IV model
\begin{equation}
\begin{aligned}
W_j & = \alpha_0 + \pi_{1}Z_j + \bX_j^\top \balp + \epsilon_{1j},\\
Y_j & = \beta_0 + \pi_{2}W_j + \bX_j^\top \bbet+ \epsilon_{2j}, \\
\end{aligned}
\label{eq:iv_model}
\end{equation}
where $\bX_j$ is the request-level covariates. $\pi_2$ is the position effect of interest. A few comments are in order. Equation \ref{eq:iv_model} imposes a linear restriction on the position effect. That is, it computes an \textit{average} position effect, such that, e.g., an increase in three positions has the same effect as three times the increase in a single position. Further, this assumption rules out position effects that vary based on what the starting position, $k_2$ is. To relate this assumption with the notation in Equation \ref{eq:poseffdef}, it imposes that $\tau_i(k_1, k_2) = c_i \times (k_2 - k_1)$, for some constant $c_i$. When the position effects have such a linear relationship, the indirect least squares (ILS) or two-stage least squares (2SLS) estimator of model \ref{eq:iv_model} with $W_j$ treated as a continuous variable provide an unbiased estimate of $c_i$: $\E[\hat\pi_2] = c_i$.

The instruments, $Z_j$ we pick are relevant because they are part of routine model tuning at LinkedIn, which operate precisely through the channel of reranking items. Finally, we assume the exclusion restriction: that the instruments only affect outcomes, $Y_j$ via their effect on positions, $W_j$. We find this assumption plausible by picking our instruments carefully, leveraging the two-sided nature of the marketplaces we study. In the ads marketplace, we have users and advertisers; in PYMK, we have two kinds of users: viewers (those who send connection invites) and candidates (those who receive connection invites). We leverage institutional knowledge in the construction on the rankings, which are determined by an ensemble AI model that includes a model that predicts user-side relevance (whether the user will click on the ad or send a connection invite) and advertiser-side relevance (how much do sponsored advertisers bid for a particular user in ads or what the predicted connection acceptance probability is in PYMK). As we seek to estimate a item-side position effect, i.e. the effect of changing only the position at which an item appears on the interaction probability of users, we must tackle confounding from user-side relevance of items. We do this by picking instruments that tune the model concerning the opposite side. Such model tuning occurs to improve relevance for sponsored advertisers in ads and the connection acceptance probability in PYMK. At the same time, they do not directly affect user-side relevance. Thus, these instruments induce exogenous variation in positions such that they do not affect outcomes, $Y_j$ except through their effect on rankings, thereby satisfying the exclusion restriction.

To estimate $\tau(k_1,k_2)$, it is intuitive to consider an estimator based on the estimated item-level position effects. For example, let $i=1,2,\ldots, N$ be a random sample of items. One can estimate $\tau(k_1,k_2)$ with 
\begin{equation}
\begin{aligned}
\hat\tau(k_1,k_2) = \frac{1}{N}\sum\limits^{N}_{i=1}\hat\tau_i(k_1,k_2),
\end{aligned}
\end{equation}
where $\tau_i(k_1,k_2)$ is the ILS or 2SLS estimator based on model (\ref{eq:iv_model}). Although this estimator is unbiased for $\tau(k_1,k_2)$, its variance can be hard to compute due to dependencies among the item-level estimators in certain applications. To see this, note that in Table \ref{tab:ex}, responses in request 1 appear in the estimate of $\tau_1(1, 2)$ and $\tau_2(1, 2)$. Since responses in the same request are likely to be related, the naive variance estimator of $\hat\tau$ can be biased. In an extreme case where there can be at most one positive response in each request (as in our ads application), this relation cannot be ignored. To deal with this structured dependency, we randomly sample one row (item) in each request and only use responses in the sampled rows to estimate $\tau$. This sampling step ensures the independence of observations in the inference for $\tau(k_1, k_2)$.

In the following sections, we provide details on how to apply this IV model in our two applications. We will go over what outcomes we estimate the position bias over, what our instrumental variables design is, which data we use, and what our empirical estimates are. In Section \ref{sec:ad-application}, we study the ads application and in Section \ref{sec:pymk-application}, we look at the PYMK application.

\section{Detailed Application: Ads Click-Through Rate}
\label{sec:ad-application}

The key goal of this application is to learn the potential change in click-through rate (CTR) when we change an ad's position in the feed. Without running an experiment, this is a very challenging question. Running an experiment, however, can potentially cause a meaningful loss in revenue. After all, ads displayed at different positions were ranked there for good reason; higher-placed ads are predicted to have higher relevance for the given search. But we observe such different rates at different positions, that it is extremely reasonable to expect that both position and underlying ad relevance determine ad clicks. Our current implementation adjusts CTR predictions for ads in positions below the top position, but without any modelling of the causal relationship between position and CTR. 

\subsection{Instrument Selection}

Experiments that affect ad position often do so exactly by changing the models that predict ad relevance. Such experiments, however, rule out using predicted relevance as a feature in our model. This is a problem, because we know that an ad's true relevance should be correlated with both position and click probability; hence our best estimates of relevance should be included as a feature in the model. Naturally, if those estimates are different under treatment and control, we cannot include them as features.

This motivates our instrument selection, a prior experiment that affected ad positions not by changing the predicted CTR (and hence the estimated ad relevance), but by experimenting with the ad bidding procedure itself, on the user level. 
We leverage an experiment that compared the existing bidding procedure, to a new reinforcement learning-based, automatic bidding method, to help advertisers determine bid values. 
Observe that this experiment has no impact on predicted relevance. 
This experiment naturally has a strong impact on ad position, but still allows us to use consistent ad relevance estimates as features in the IV model. 

\subsection{Data Collection and Sampling}

Observe that an instrumental variables model works on the relevance of the instrument, i.e. on the nonzero effect of the instrument on the treatment, in this case the feed position. If we consider the full set of ad campaigns (a campaign is the ads from a specific advertiser) together, this regression is not meaningful. For a given user and query, several ads will appear; and moreover, if ad A is in position 1, ad B cannot be in position 1, because it is already occupied. This introduces problematic correlation. Consider the toy example in which we have only two ads, A and B. Then the position of A exactly determines B and our regression coefficients are not identifiable. While in our case we likely could find regression coefficients, they are not trustworthy due to this collinearity. Moreover, if we were attempting to perform a stage 1 regression, the coefficient is essentially meaningless when we consider the full set of ads in one regression. 

We therefore assign the level of estimation for this regression to be each ad campaign. By reducing the data to those requests featuring a specific ad campaign, we only sample one observation per request, since the same ad campaign does not typically appear multiple times in a single request. Thus, by only sampling at most a single observation per request, we remove problematic correlation across observations in the same request. For this analysis, we considered the 100 largest ad campaigns, which amounts to 44,349,019 ad views across the top 30 feed positions on one day. There are two treatment variants, randomized at the user level. 

\subsection{Model Specifications}

We consider three model specifications to estimate the position effects (Table \ref{table:modelspecs_ads}). Specification 1 is the simplest IV specification, which seeks the effect of feed position on 1(click) using the treatment as our instrument. Specification 2 is Specification 1 with the addition of a control variable, predicted CTR (PCTR), attempting to capture ad relevance. Recall that the ad relevance model remains stable throughout the duration of the experimentation we use as instruments.  Specification 3 is an ordinary least squares (OLS) non-causal baseline, which does not include the instrument (Experiment = treatment). We prefer Specification 2; this allows us to leverage the past randomness and estimate a causal effect, while also controlling for any residual effects of predicted ad relevance in the model. 

\begin{table}[h]
\centering 
\resizebox{\textwidth}{!}{
\begin{tabular}{lccc}
\hline \hline 
 & Spec 1 (IV) & Spec 2 (IV, preferred) & Spec 3 (OLS) \\
 \hline 
 Outcome & 1(click) & 1(click) & 1(click) \\
Endogenous variables & Feed position & Feed position & N/A \\
Instruments & Experiment = treatment & Experiment = treatment &  N/A \\
Control variables & N/A & PCTR & Feed position, PCTR \\
 \hline \hline 
\end{tabular} }
\caption{Regression model specifications in the ads analysis.}\label{table:modelspecs_ads}
\end{table}

\subsection{Results}

The most robust IV results will correspond to a strong stage 1 regression, meaning a low p-value and a relatively large coefficient; otherwise, we are unlikely to be able to detect a causal effect. Moreover, it is reasonable to assume that the bidding experiment will affect different ad campaigns differently, and indeed we have a range of stage 1 regression information. 

\begin{figure}[h]
    \centering
    \includegraphics[width=\textwidth]{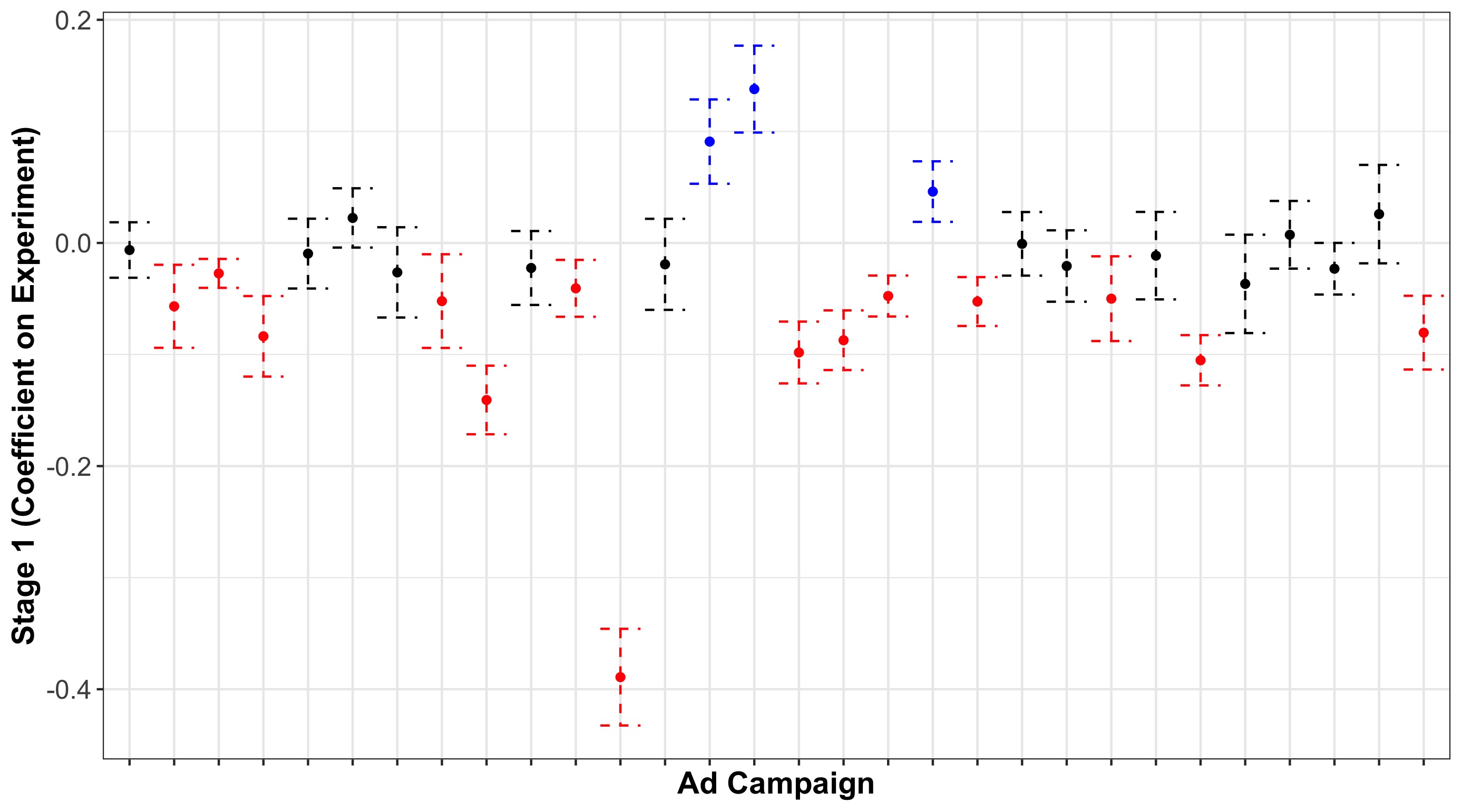}
    \caption{Coefficient showing the impact of the bidding experiment, which was randomized on the user level, on feed position for individual ad campaigns. Observe that these coefficients correspond exactly to the first stage of our instrumental variables model. Here we show the 30 ad campaigns with the most feed views in our time window.}
    \label{fig:stage1_coefficients}
\end{figure}

Figure \ref{fig:stage1_coefficients} shows the coefficient on the bidding experiment's effect on feed position, for the 30 ads with the most feed views, along with 95\% confidence intervals. A negative coefficient here indicates that the bidding experiment used as the instrument on average moved ads up in the feed (decreased their feed position). Results are color-coded according to their distance from zero; campaigns with a significantly negative coefficient in red, positive coefficient in blue, and not different from zero in black. While the plurality of ad campaigns (47\%) have a negative impact, we see many without a significant impact (43\%), and a few with a positive impact (10\%). 

\begin{figure}[h]
    \centering
    \includegraphics[width=0.8\textwidth]{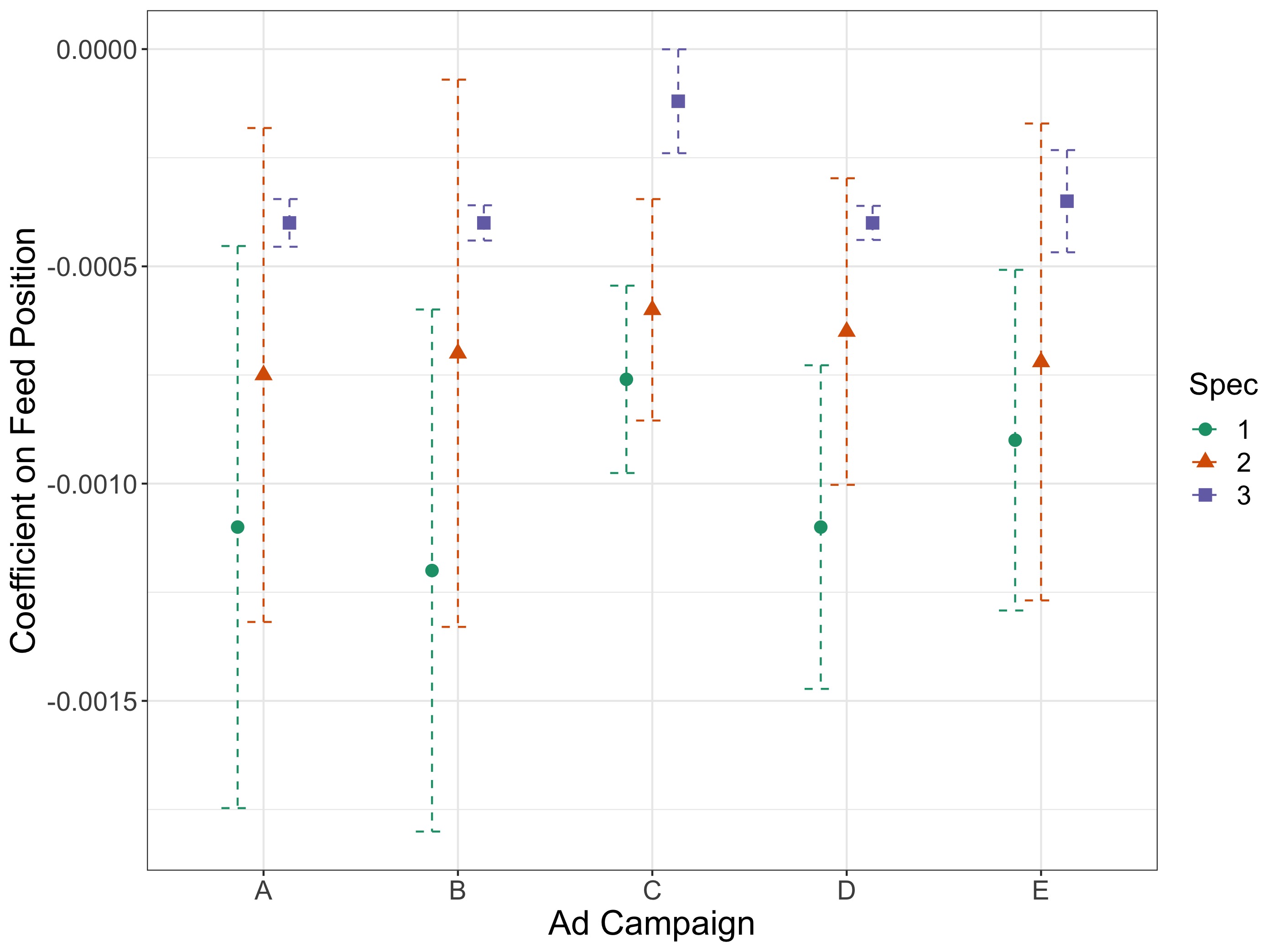}
    \caption{Estimates of position bias effect from model specifications 1, 2, and 3 for five ad campaigns.}
    \label{fig:top_five_ads}
\end{figure}

For Figure \ref{fig:top_five_ads}, we select the ads with the most statistically significant stage 1 coefficients among the 50 ads with the most feed appearances. 
This gives us an estimate of position bias for campaigns with the most robust available IV estimation. 

The position effect here is our coefficient on feed position.
These results show a position effect across several model specifications, including a model accounting for predicted click-through rates (our preferred model specification). We see these results for ad campaigns in which our instrument, namely the bidding experiment, had a strong impact on the feed position in which the ad appeared. 
Hence interpretation should be constrained as follows: in ad campaigns that were strongly affected by the bidding experiment, there is a robust and statistically significant effect of ad position on click-through rate. These ad campaigns might differ systematically from the campaigns without as large an impact from the bidding experiment, and that difference could apply to position effects as well. 

For more details, in Table \ref{table:adsregression} we display the full regression information for ad campaign A from Figure \ref{fig:top_five_ads}. Across the model specifications, moving an ad to a higher feed position decreases click probability; higher PCTR is consistently correlated with greater click probability. In model (2), the coefficient estimating the impact of feed position drops in magnitude. This is likely because we include PCTR in this model, and the feed position coefficient in model (1) accounts for lower feed position being correlated with higher PCTR.

The coefficients on feed position are consistently negative, which lines up with intuition that moving an ad up higher in the feed (that is, decreasing feed position) will raise click probability. For our preferred specification, the coefficient for ad campaign A on feed position is -0.0007 (with a standard deviation of 0.0003).
That is, if we move ad A from position $k$ to position $k-1$, the probability of a click will increase on average (over users) by $+0.07\%$. 
Compared to the constant estimate of $0.58\%$, this amounts to an increase of $12\%$. 
\citet*{rddlinkedin} also consider position bias for ads, comparing specifically CTR in the top two ad positions with a regression discontinuity model where the running variable is the difference in relevance score predictions for a pair of adjacently positioned ads. They find that the effect of going from the lower to the higher position on CTR is an increase in probability of $50.81\%$. This discrepancy is intuitive; the regression discontinuity model gives an estimate for a pair of ads where the relevance scores are exactly equal, and hence CTR is highly likely to be affected by position. The instrumental variables model includes relevance as a predictor, but gives an estimate across ads with a range of predicted relevance scores, and therefore should be expected to give an estimate with lower magnitude.

\begin{table}[H] \centering 
  \resizebox{\textwidth}{!}{
\begin{tabular}{@{\extracolsep{5pt}}lccc} 
\\[-1.8ex]\hline 
\hline \\[-1.8ex] 
 & \multicolumn{3}{c}{\textit{Dependent variable:}} \\ 
\cline{2-4} 
\\[-1.8ex] & \multicolumn{3}{c}{Ad click (1 or 0)} \\ 
\\[-1.8ex] & IV & IV & OLS\\ 
\\[-1.8ex] & (1) & (2) & (3)\\ 
\hline \\[-1.8ex] 
Feed position & $-$0.0012$^{***}$ & $-$0.0007$^{**}$ & $-$0.0004$^{***}$ \\ 
  & (0.0003) & (0.0003) & (2.07$\times10^{-5}$) \\ 
  & & &  \\ 
PCTR &  & 0.70$^{***}$ & 0.72$^{***}$ \\ 
  &   & (0.032) & (0.024) \\ 
  & & &  \\ 
Constant & 0.016$^{***}$ & 0.0058$^{**}$ & 0.0035$^{***}$ \\ 
  & (0.003) & (0.002) & (0.0003) \\ 
  & & &  \\ 
\hline \\[-1.8ex] 
Observations & 402,358 & 402,358  & 402,358  \\ 
Residual Std. Error & 0.089 (df = 402358) & 0.089 (df =  402357) & 0.089 (df =  402357) \\ 
\hline 
\hline \\[-1.8ex] 
 & \multicolumn{3}{r}{$^{*}$p$<$0.1; $^{**}$p$<$0.05; $^{***}$p$<$0.01} \\ 
\end{tabular} }
\caption{Detailed regression information for ad campaign A across all model specifications.}\label{table:adsregression}
\end{table} 

\section{Detailed Application: People-You-May-Know Invitations}
\label{sec:pymk-application}

Every week, millions of LinkedIn users visit the PYMK page and view recommendations on who to connect with on the platform. The recommendations are chosen and ranked according to an ensemble model that aims to maximize the probability of a connection made between the viewer and the candidate recommended and interactions between them. 

Given this, we seek to identify the position effect of invites sent in PYMK. That is, conditional on the match quality between two users, the question we ask is: what is the additional likelihood that the viewer will send an invitation to the candidate just because of their position on the PYMK rankings? In other words, in a given session, if the same candidate was ranked in position 2 as opposed to position 3, would they see a higher probability of receiving a connection invite? 

We frame this as a causal inference problem because we want to isolate the effect of changing only position. A correlational analysis of candidate positions and the subsequent probability of invites received that they realize is confounded by the fact that higher-ranked candidates, on average, are more relevant to the viewers (in the sense of actually leading to a successful connection formed). We thus understand that a causal position effect would identify the behavioral response on the part of viewers to send connection requests on PYMK, holding fixed candidate quality or relevance.

\subsection{Context}
The PYMK offline ranker is an ensemble model consisting of four constituent AI models: 1) pInvite: a model predicting the probability that the viewer sends the candidate an invite, 2) pAccept: a model predicting the probability that the candidate accepts an invite from the viewer, 3) destSessionUtility: a model predicting the interaction between the two once connected, and 4) destRetentionUtility: a model measuring broader engagement of the candidate on linkedin.com after connecting with the viewer. Candidates are ranked based on a multi-objective optimization function (MOO), which is a function of the predicted values of each one of these four models. 

In particular, the pInvite model is trained using historical PYMK data, i.e., candidates who have been impressed in viewers' PYMK and their corresponding outcomes (whether the viewer sent an invite) become labels for training. Ignoring position effect during model training can potentially yield biased estimates if two candidates' labels are treated equally by the model but one candidate was positioned lower than the other.

\subsection{Instrument Selection}
Our proposal for causal identification is to use A/B testing in the constituent AI models of the PYMK MOO score. Specifically, we look at experimentation in the pAccept model and the weights of each of the individual AI models in the overall MOO score. 

Since the experiments are randomized on viewers, any systematic difference we see in the positions candidates take between the PYMK grids of viewers in the treatment group and those in the control group is random and therefore unrelated to viewer characteristics.  

Thus, we may use the PYMK AI experiments as instruments to exogenously move the position of candidates, which we may then use to identify the position effect on the probability of sending PYMK connection invites. 

\subsection{Data Collection and Sampling}
Our dataset is at the edge-level, meaning that each observation corresponds to one (viewer, candidate, query) tuple. It consists of 38,553,973 observations, 3,823,649 unique viewers, 24,596,318 unique candidates, and 22 PYMK related reasons. There are two treatment variants, randomized at the viewer level. We also observe pInviteScore: a PYMK AI model that indicates the likelihood that the viewer will send an invite to the candidate, as well as pInviteOutcome: the final outcome of whether the viewer sent a connection invite to the candidate or not. Finally, we observe the rank each candidate was displayed to the viewer in the latter’s PYMK query, as well as the session depth, i.e. the number of recommendations the viewer was served in a particular session. 

\subsection{Model Specifications}
The independent variable we seek to model is pInviteOutcome. We use the instrumental variables (IV) framework for our causal results here. To identify multiple endogenous variables, we require multiple instruments. Given we have a single binary treatment, we interact it with relatedReason to give us some two dozen instruments. 

Given this setup, we run the following edge-level regression specifications: 
\begin{table}[H]
\begin{tabular}{lcccc}
\hline \hline 
                     & Spec 4 (IV)                                                          & Spec 5 (IV)                                                          & Spec 6 (IV)                                                          & Spec 7 (OLS)                                                                         \\ \hline 
Outcome              & pInviteOutcome                                                       & pInviteOutcome                                                       & pInviteOutcome                                                       & pInviteOutcome                                                                       \\
Endogenous variables & onlineMooRank                                                        & onlineMooRank                                                        & \begin{tabular}[c]{@{}l@{}}onlineMooRank, \\ sessDepth\end{tabular}  & N/A                                                                                  \\
Instruments          & \begin{tabular}[c]{@{}l@{}}treatment $\times$ \\ relatedReason\end{tabular} & \begin{tabular}[c]{@{}l@{}}treatment $\times$ \\ relatedReason\end{tabular} & \begin{tabular}[c]{@{}l@{}}treatment $\times$ \\ relatedReason\end{tabular} & N/A                                                                                  \\
Control variables    & No                                                                  & pInviteScore                                                         & pInviteScore                                                         & \begin{tabular}[c]{@{}l@{}}onlineMooRank, \\ sessDepth, \\ pInviteScore 

\end{tabular} \\
\hline \hline 

\end{tabular}
\caption{Edge-level regression model specifications in the PYMK analysis.}
\label{table:pymk-edge-modelspecs}
\end{table}

Specification 4 is the simplest IV specification that seeks the effect of onlineMooRank on pInviteOutcome. Specification 5 is Specification 4 with the addition of the control variable, pInviteScore. Since our treatment variants did not explicitly seek to randomize PYMK positions between treatment and control, there may be systematic match quality differences between the two variants. For this reason, pInviteScore acts as a way to control for match quality. 

Our preferred specification is 6, which is Specification 5 with an additional endogenous variable, sessDepth. Again, the treatment did not just randomize PYMK position. If session depths vary systematically between treatment and control, that may indicate a difference in match quality, and this difference should not be driving our IV results. As sessDepth responds to treatment, we include it as an endogenous variable in this specification. Finally, as a baseline, we include ordinary least squares (OLS) estimates that do not involve any instruments.  

\subsection{Results}
\begin{table}[H] \centering 
  \resizebox{\textwidth}{!}{
\begin{tabular}{@{\extracolsep{5pt}}lcccc} 
\\[-1.8ex]\hline 
\hline \\[-1.8ex] 
 & \multicolumn{4}{c}{\textit{Dependent variable:}} \\ 
\cline{2-5} 
\\[-1.8ex] & \multicolumn{4}{c}{pInviteOutcome} \\ 
\\[-1.8ex] & IV & IV & IV & OLS\\ 
\\[-1.8ex] & (4) & (5) & (5) & (7)\\ 
\hline \\[-1.8ex] 
 onlineMooRank & $-$0.038$^{***}$ & $-$0.038$^{***}$ & $-$0.063$^{***}$ & $-$0.005$^{***}$ \\ 
  & (0.001) & (0.0005) & (0.001) & (0.0001) \\ 
  & & & & \\ 
 sessDepth &  &  & 0.013$^{***}$ & 0.022$^{***}$ \\ 
  &  &  & (0.001) & (0.0001) \\ 
  & & & & \\ 
 Constant & 0.192$^{***}$ & 0.199$^{***}$ & 0.210$^{***}$ & $-$0.196$^{***}$ \\ 
  & (0.003) & (0.003) & (0.003) & (0.001) \\ 
  & & & & \\ 
  Controls & No & pInviteScore & pInviteScore & pInviteScore \\
  & & & & \\
\hline \\[-1.8ex] 
Observations & 3,855,397 & 3,855,397 & 3,855,397 & 3,855,397 \\ 
R$^{2}$ & $-$0.520 & $-$0.513 & $-$0.822 & 0.158 \\ 
Adjusted R$^{2}$ & $-$0.520 & $-$0.513 & $-$0.822 & 0.158 \\ 
Residual Std. Error & 0.231 (df = 3855395) & 0.230 (df = 3855394) & 0.253 (df = 3855393) & 0.172 (df = 3855393) \\ 
\hline 
\hline \\[-1.8ex] 
\textit{Note:}  & \multicolumn{4}{r}{$^{*}$p$<$0.1; $^{**}$p$<$0.05; $^{***}$p$<$0.01} \\ 
\end{tabular} }
  \caption{Estimates from various regression specifications in the edge-level PYMK analysis.} 
  \label{table:estimates-edge-pymk} 
{\small Standard errors are clustered at the viewer level, at which level the treatment is randomized.} 
\end{table} 

The coefficient on onlineMooRank is our position effect of interest. We find consistently non-zero position effects across our regressions, with our IV point estimates spanning the range between -0.063 and -0.038. In words, this means that when a candidate is bumped up one rank on the PYMK page, then on average they are likely to see between 3.8\% and 6.3\% higher probability of receiving a connection invite.

\section{Discussion}
\label{discus}
In this work, we introduce new methodology to derive causal position effects from observational data by leveraging marketplace instrumental variables. We find variation in the rankings of items that is exogenous to user relevance in experimentation on the item side of the marketplace. We validate the methodology on two distinct domains---the PYMK connections recommender and the ads marketplace---and are able to measure position effects in a range of contexts. The existence of position effects is important to note in model training, where impressed items are matched with their realized outcomes (labels) to form the training data. The position effects we estimated stipulates that lower-ranked items would have realized a better outcome had they been positioned higher, even conditional on impression. Therefore, to train a model without adjusting for position effects means that we are understating the labels of lower-ranked items relative to those higher-ranked items in model training, which could lead to model bias if lower-ranked items are systematically different than higher ranked items in the features that are included in the model. Thus, our estimates can be used in ranking model retraining for debiasing, such as by appropriately discounting user action for the item they interacted with by the modeled position bias in future training data. Such a correction plays an important role in the user action calibration layer of recommender systems. 

Our empirical results in PYMK show a robust position effect across multiple specifications. We thus have evidence for a non-zero and economically significant position effect among users when sending out connection invitations. This position effect is after accounting for match quality, and is thus a behavioral artifact on the part of users. Analogously, we see a consistent position effect for ads in the feed, observed across a range of ad campaigns highly impacted by the bidding experiment (that is, with a strong instrument). In addition to demonstrating the robustness of our empirical estimates, we would like to make a few points regarding how to interpret the result in the context of the PYMK UI. First, the multiple instruments we used were by interacting the treatment, which was randomized across users, with the relatedReason, which was not. Thus our estimates represent a homogeneous position effect across all the relatedReason sections on the PYMK, each of which is presumed to have similar effects as the others. In future studies, we can relax this assumption by using more experimental variants as instruments. Second, we have not dealt with interference and SUTVA violations in this analysis. While we did cluster standard errors at the viewer level, that accounts for correlations within a viewers but not those across candidates. These spillovers may occur when, for instance, a viewer in the treatment variant sends an invite to one in a control variant, thereby preventing the latter from sending an invite to the viewer. Such spillovers can be very complex to model and are usually not addressed in PYMK experiments. We leave the exercise of teasing apart spillover effects from principal effects to future work. 

In contrast, for the ads example, we are able to manage interference by looking at each campaign separately; a user does not usually see more than one ad from a given campaign in her feed at any given time. This comes along with its own limitations on interpretation; namely, for ad campaigns that did not see major ranking changes due to the bidding experiment, we are unable to give a robust position effect estimate. Our results should be constrained as estimates for those that were significantly effected by our choice of instrument. These ad campaigns might differ systematically from the campaigns without as large an impact from the bidding experiment, and that difference could apply to position effects as well. Moreover, there is selection bias in our sampling scheme, due to which users were served a given campaign and which were not. This is inherent in our data collection scheme, but should constrain interpretation. 

Future research should consider how to use these insights from observational causal analyses to design experiments to complement these learnings. In the ads example, we identified a position effect among ad campaigns that had ranking changes due to the bidding experiment; we might consider selecting a subset of the remaining ads as strong candidates for randomization, following for example \citet*{RosenmanOwen}. Another direction for future work is to explore differences between ad campaigns or LinkedIn users. It is likely that the effect of position is heterogeneous across those groups, and that position bias modelling would be improved greatly by understanding the complex heterogeneous treatment effects present.

\bibliographystyle{apalike}
\bibliography{biblio}

\newpage
\appendix

\section{Session-level model}
Given that the PYMK experiment we use is randomized at the user level, a specification that is closer to the user level might be desired. In this section, we present results at the session level. We define “top spot” as candidates being in the top 4 positions of the PYMK recommendation page and “bottom spot” when they are in positions 5 or higher. We then aggregate how many candidates were shown in each session in top spots (nTopSpot) and in the bottom spots (nBottomSpot), as well as how many total connection invites were sent by the viewer in that particular session (pInviteOutcomeTotal). As in the edge-level specifications, we use our PYMK experiment interacted with relatedReason as our instruments. 

Given this setup, we run the following session-level regression specifications: 
\begin{table}[H]
\centering
\begin{tabular}{lcc}
\hline \hline 
                     & Spec A1 (IV)              & Spec A2 (OLS baseline) \\ \hline 
Outcome              & pInviteOutcomeTotal       & pInviteOutcomeTotal    \\
Endogenous variables & nTopSpot, nBottomSpot     & N/A                    \\
Instruments          & treatment $\times$ relatedReason & N/A                    \\
Control variables    & No                       & nTopSpot, nBottomSpot \\
\hline \hline 
\end{tabular}
\caption{Session-level regression model specifications in the PYMK analysis.}
\label{table:pymk-sess-modelspecs}
\end{table}

Our preferred specification is Specification A1, the IV specification. We include no control variables because aggregating pInviteScore across candidates in a session may lead to a variable that is meaningless to compare across sessions. 

\subsection{Regression estimates}
\begin{table}[H] \centering 
\begin{tabular}{@{\extracolsep{5pt}}lcc} 
\\[-1.8ex]\hline 
\hline \\[-1.8ex] 
 & \multicolumn{2}{c}{\textit{Dependent variable:}} \\ 
\cline{2-3} 
\\[-1.8ex] & \multicolumn{2}{c}{pInviteOutcomeTotal} \\ 
\\[-1.8ex] & IV & OLS \\ 
\\[-1.8ex] & (1) & (2)\\ 
\\[-1.8ex] &  & \\ 
\hline \\[-1.8ex] 

 nTopSpot & 0.015$^{**}$ & $-$0.009$^{***}$ \\ 
  & (0.007) & (0.001) \\ 
  & & \\ 
 nBottomSpot & $-$0.130$^{***}$ & 0.208$^{***}$ \\ 
  & (0.002) & (0.001) \\ 
  & & \\ 
 Constant & 0.394$^{***}$ & $-$0.169$^{***}$ \\ 
  & (0.025) & (0.002) \\ 
  & & \\ 
 Controls & No & No \\
  & & \\
\hline \\[-1.8ex] 
Observations & 7,083,918 & 7,083,918 \\ 
R$^{2}$ & $-$0.276 & 0.171 \\ 
Adjusted R$^{2}$ & $-$0.276 & 0.171 \\ 
Residual Std. Error (df = 7083915) & 1.513 & 1.220 \\ 
\hline 
\hline \\[-1.8ex] 
\textit{Note:}  & \multicolumn{2}{r}{$^{*}$p$<$0.1; $^{**}$p$<$0.05; $^{***}$p$<$0.01} \\ 
\end{tabular} 
  \caption{Estimates from regression specifications in the session-level PYMK analysis.} 
  \label{table:estimates-sess-pymk} 
{
\small Standard errors are clustered at the viewer level, at which level the treatment is randomized. }

\end{table} 

Our session-level IV results show an average position effect from a bottom spot to a top spot is 0.145 (the difference between the coefficients on nTopSpot and nBottomSpot). In words, this says that on average, when a candidate is bumped up from a bottom spot (position 5 or larger) to a top spot (position 4 or smaller), they receive an extra 14.5\% probability of receiving a PYMK connection invite. The magnitude of this estimate is in the same order as the single-position-incrementing estimate obtained from the edge level regressions, which is reassuring.

Importantly, the OLS regression results are not immediately rationalizable. They indicate a position effect of the opposite sign, i.e. indicating that candidates at the bottom of the PYMK page might receive more invites. Such a result that does not pass a basic sanity check would require more exacting scrutiny to hold true. We take the stance that the OLS result is overwhelmed by confounding to present a valid position effect in the way we understand it. 

We conclude this section with a critical examination of various levels of aggregating for a study such as this. 

\subsection{A comparison of different levels of aggregation}
\begin{table}[H]
\centering

\resizebox{\textwidth}{!}{
\begin{tabular}{p{0.2\linewidth}  p{0.4\linewidth}  p{0.4\linewidth}}
\hline \hline 
& Pros & Cons \\ \hline 
Edge level & 
\begin{enumerate}
    \item Uses the raw data, as collected.
    \item Includes explicit relationship between which viewer sent an invite to which candidate.
    \item Can obtain fine grained relationship between exact position effect and outcome.
\end{enumerate} & 
\begin{enumerate}
    \item PYMK experiment is not randomized at the edge level.
\item If viewed as a cluster-randomized experiment, cluster size is endogenous to experiment. 
\end{enumerate} \\ \hline 
Session level & 
 The level at which positions are assigned. & 
\begin{enumerate}
    \item PYMK experiment not randomized at the session level.
    \item If viewed as a cluster-randomized experiment, cluster size (i.e. number of sessions per viewer) is endogenous to experiment. 
    \item Need to aggregate outcome at session level, losing information on candidates. Also means controlling for match quality (via pInviteScore) is harder.
    \item Need to aggregate position data at the session level, making choice of endogenous variable a subject of potentially arbitrary design decisions. 
\end{enumerate} \\ \hline 
user (viewer) level & 
The same level as at which the experiment was randomized. (Cleanest specification from the point of view of experiment design.) & 
\begin{enumerate}
    \item Same aggregation issues as session level.
\item On top of that, we are now aggregating data across sessions, potentially threatening identification of position effect. (Positions are defined at the session level, so aggregating positions across different sessions is problematic.)
\end{enumerate} \\
\hline \hline 
                                                                                                         
\end{tabular} }
\caption{A critical look at PYMK analysis performed at various levels of data aggregation: the edge level, where each observation corresponds to one viewer and candidate pair from a particular session; the session level, where observations corresponding to a particular session are aggregated into a single observation; and the viewer level, which aggregates all sessions from a given viewer (i.e. sender of connection invites) into one observation.}
\label{table:pymk-aggregation-comparison}

\end{table}

\end{document}